\documentclass[aps,prb,twocolumn,superscriptaddress,showpacs,amsmath,amssymb,
amsfonts,balancelastpage]{revtex4-1}
\usepackage{amsmath}
\usepackage{amssymb}
\usepackage{amsfonts}
\usepackage[justification=raggedright]{caption}
\usepackage{subcaption}
\usepackage{graphicx}
\usepackage{epsfig}
\usepackage{hyperref}
\usepackage{epstopdf}
\usepackage{gensymb}
\usepackage{color }
\usepackage{color, soul }
\usepackage{xcolor}
\hypersetup{
	colorlinks=true,
	urlcolor=blue,
	citecolor=blue,
	linkcolor=blue,}
\begin{document}
		\title{ Weak Anti-localization in thin films of the Topological Semimetal Candidate Pd$_{3}$Bi$_{2}$S$_{2}$}
	
	\author{Shama}
	\affiliation{Department of Physical Sciences, Indian Institute of Science Education and Research, Knowledge city, Sector 81, SAS Nagar, Manauli PO 140306, Mohali, Punjab, India}
	
	\author{Goutam Sheet}
	\affiliation{Department of Physical Sciences, Indian Institute of Science Education and Research, Knowledge city, Sector 81, SAS Nagar, Manauli PO 140306, Mohali, Punjab, India}
	
	\author{Yogesh Singh} \email{yogesh@iisermohali.ac.in}
	\affiliation{Department of Physical Sciences, Indian Institute of Science Education and Research, Knowledge city, Sector 81, SAS Nagar, Manauli PO 140306, Mohali, Punjab, India}
	
\begin{abstract}
We report the growth and magneto-transport studies of Pd$_{3}$Bi$_{2}$S$_{2}$ (PBS) thin films synthesized by pulsed laser deposition (PLD) technique. The magneto-transport study on pristine and post annealed films show the presence of more than one type of charge carrier with a carrier concentration in the range $0.6$ - $2.26~\times$ 10$^{21}$ cm$^{-3}$ and mobility in the range 0.96 - 1.73 $\times$ 10$^{2}$ cm$^{2}$/Vs.  At low temperatures a logarithmic increase in conductivity is observed which indicates the presence of weak anti-localization (WAL).  The magnetotransport data is analysed within the Hikami-Larkin-Nagaoka (HLN) theory.  It is found that temperature dependence of the dephasing length can't be explained only by electron-electron scattering and that electron-phonon scattering also contributes to the phase relaxation mechanism in PBS films.
\end{abstract}
\maketitle

\section{Introduction}
Topological materials such as topological insulators, topological semimetals, etc. have gained wide interest in condensed matter physics owing to their novel physical properties and potential applications in the field of spintronics. \cite{Hasan,Moore,Burkov,Wan,Armitage}  Topological Insulators (TI) have a bulk band gap with conducting edge or surface states protected by time-reversal symmetry (TRS).\cite{Hasan,Moore}  Topological semimetals (TSM) can be classified into Dirac semimetals and Weyl semimetals.\cite{Armitage,Burkov,Wan}  In a Dirac semimetal, conduction and valence band touch each other linearly at discrete points in the Brillouin zone called Dirac nodes which are protected by time-reversal symmetry (TRS) and Inversion symmetry (IS).\cite{Armitage,Young}  On breaking either TRS or IS, this Dirac node splits into two Weyl nodes with opposite chiral charge, forming a Weyl semimetal.\cite{Armitage} The quasiparticles in Weyl and Dirac semimetals have two and four-fold degeneracy. Among the TSM candidates, Cd$_{3}$As$_{2}$ and Na$_{3}$Bi were first theoretically predicted and experimentally verified to host Dirac fermions,\cite{Wang,Weng,Liu,Zhou} whereas Weyl fermion were identified in the $TX (T =$ Ta, Nb; $X =$ As, P) family of materials.\cite{Lv,Xu,Yuan} The topological semimetals exhibit unusual phenomena such as extremely large magneto-resistance, chiral anomaly induced negative MR, high mobility, and anomalous Hall effect.\cite{Zhang,Xiong,Shekhar,Liang}
    
Recently, based on certain space group symmetries, Bradlyn.\textit{et al} have theoretically predicted the existence of exotic fermions (beyond Dirac, Weyl, Majorana) having three, six, and eightfold degenerate band crossing in various materials.\cite{Bradlyn}  Pd$_{3}$Bi$_{2}$S$_{2}$ with space group 199 is one such material proposed to host exotic fermions having three-fold degenerate band crossing just $0.1$~eV away from the Fermi level.\cite{Bradlyn}  In previous studies, the magneto-transport has been reported on polycrystalline, single-crystalline, and nanoparticle samples of Pd$_{3}$Bi$_{2}$S$_{2}$. \cite{Takeshi,Maria,Roy}  The magneto-transport study on single-crystal indeed revealed interesting features like a large non-saturating magneto-resistance, and presence of more than one type of charge carriers with very high mobility. \cite{Roy}

Reduction of system dimensionality from three dimensional (3D) to two dimensional (2D), is expected to lead to the emergence of several quantum phenomena at low temperatures, including weak localization (WL) or weak anti-localization (WAL), quantum interference (QI \cite{Bergmann,Hikami,Fang,Lee}, and universal conductance fluctuations (UCF).\cite{Stone} The constructive or destructive interference of electrons moving through time reversed paths gives rise to negative or positive correction to conductivity, which is known as  weak localization or anti-localization (WL/WAL).\cite{Bergmann,Hikami,Fang,Lee}  These effects can be experimentally verified using electron transport such as magneto-resistance (MR), Hall effect, etc. WAL is observed in topological materials, as an important consequence of spin momentum locking, resulting in a relative $\pi$ Berry phase acquired by electrons executing time-reversed paths.\cite{Lu,He}

In this work, we report the first growth and characterization of Pd$_{3}$Bi$_{2}$S$_{2}$ (PBS) thin films using pulse laser deposition technique (PLD) and magneto-transport properties of PBS films prepared under different post annealing conditions.  In the magnetic field dependent conductivity at low temperature, we have found a sharp cusp around $B = 0$ when the magnetic field is perpendicular to the film, revealing the presence of weak anti-localization.  Applying the HLN theory, we have extracted the dephasing length (L$_{\phi}$). Additionally we find that the temperature dependence of L$_{\phi}$ can not be described only by an electron-electron dephasing  mechanism.\cite{Wu}  This indicates the existence of some other phase relaxation mechanism such as electron-phonon scattering in Pd$_{3}$Bi$_{2}$S$_{2}$ films.\cite{Bird,Sergeev,Reizer}
    
    \begin{figure}
\includegraphics[width=\linewidth, height=0.25\textheight]{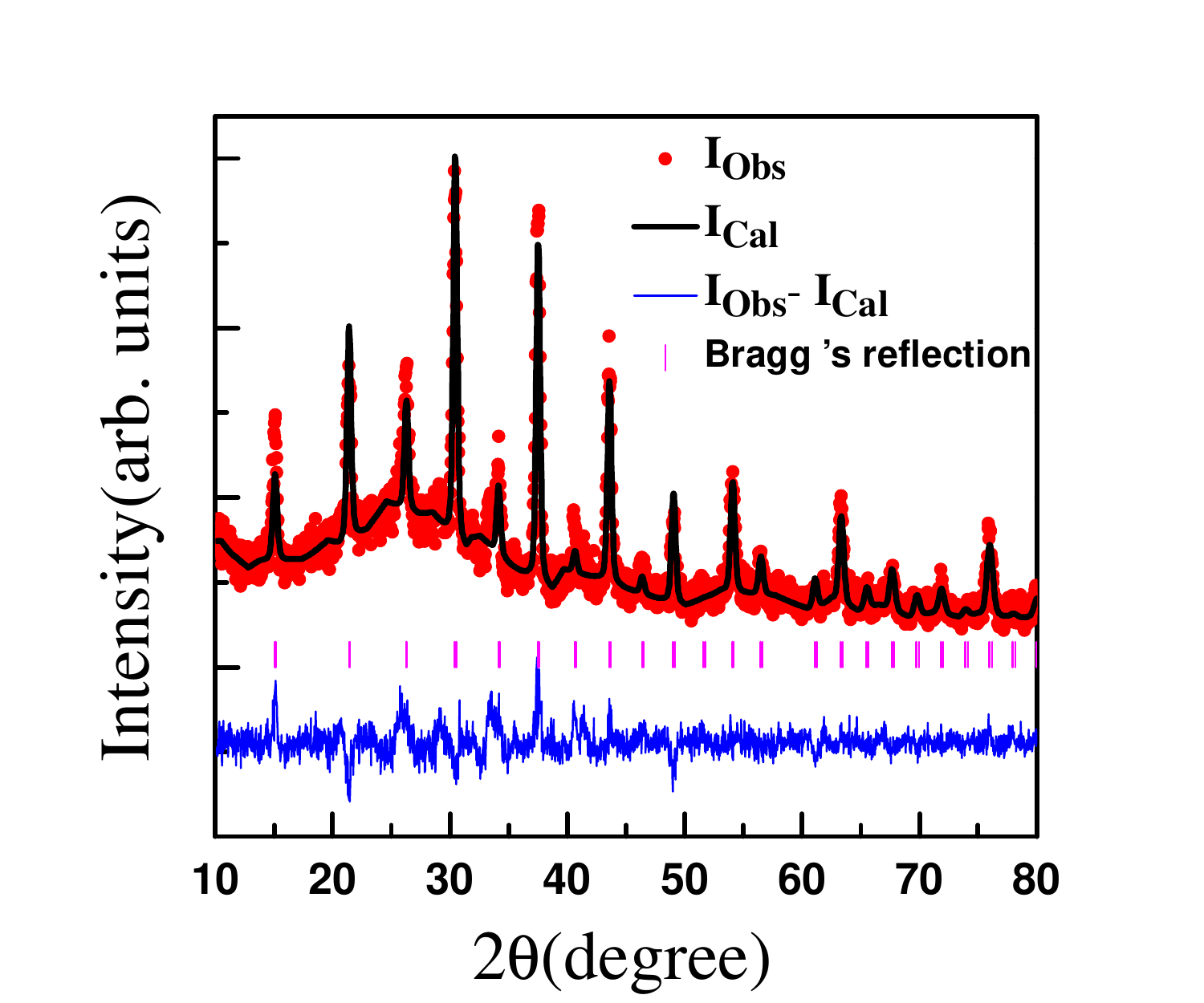}
\caption{X-ray diffraction data (solid symbols) for as-grown (S0) Pd$_{3}$Bi$_{2}$S$_{2}$ film and it's Rietveld refinement (solid curves through the data).  The positions of the expected Bragg peaks are given as the vertical bars.  The diference between the data and the refined pattern is shown at the bottom.}
\label{fig-XRD}
\end{figure}

\begin{figure}
	\includegraphics[width=1.1\linewidth, height=0.35\textheight]{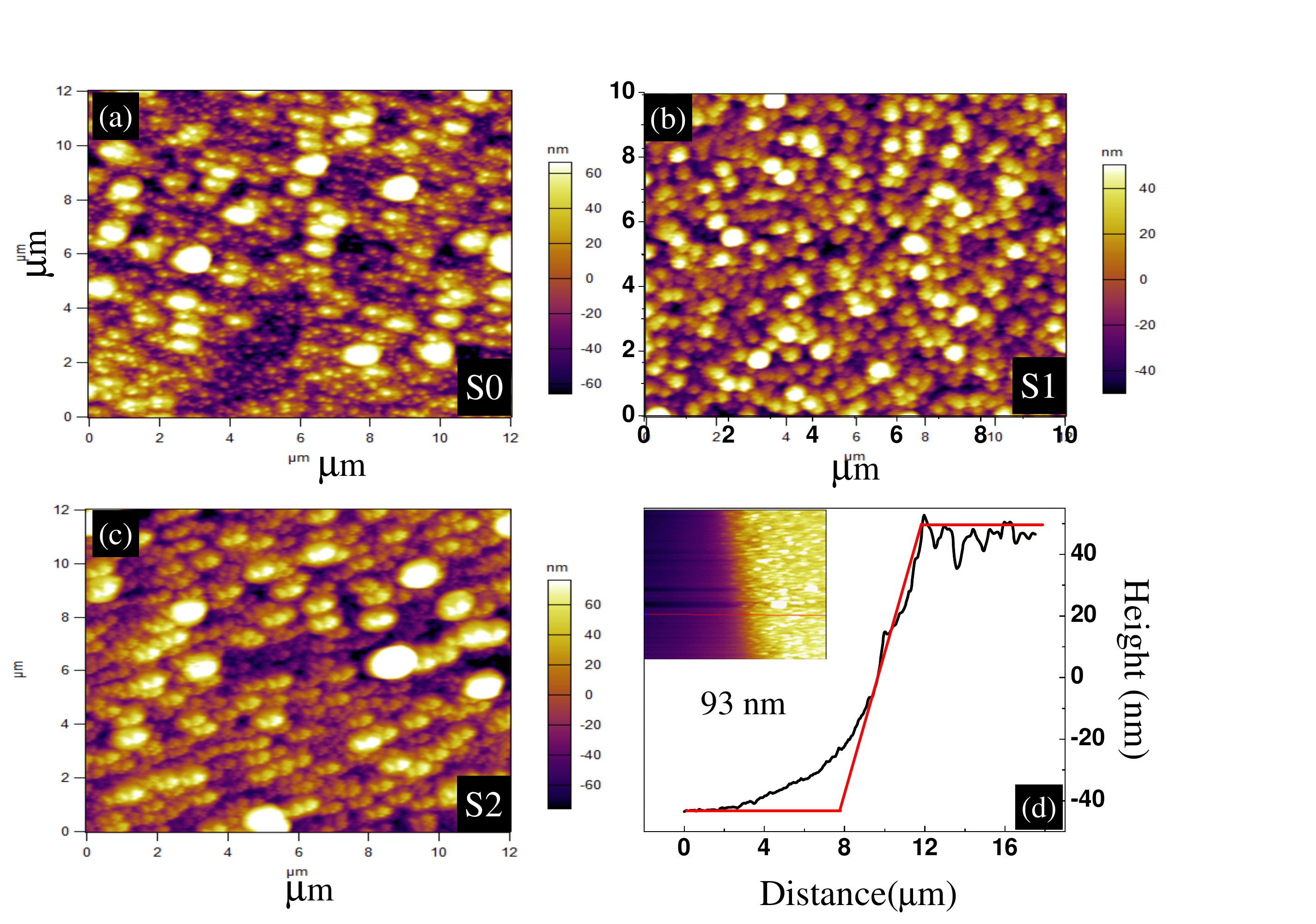}
	\caption{(a-c) Atomic Force Microscope topography image of the surface of the three PBS thin film used in this study. (d) The height profile across the S0 film with a thickness $\approx 93$~nm.}
	\label{fig-afm}
\end{figure}

\section{Methods}
Pd$_{3}$Bi$_{2}$S$_{2}$ thin films were grown on a Si(111) substrate using pulsed laser deposition ( KrF excimer , $\lambda$ = 248 nm) in Argon atmosphere. A polycrystalline target was prepared by the solid state reaction of stoichiometric amounts of high purity starting materials in a vacuum sealed quartz tube. The laser ablation was performed on the polycrystalline target under the growth conditions of 1.8 J/cm$^{2}$ laser fluence at low repetition rate of $1$Hz.\cite{Yan} The as-grown PBS thin-films (S0) were post annealed for $30$~minutes in Argon atmosphere at $260~^o$C (S1) and $300~^o$C (S2).  The x-ray diffractometer (Bruker D8 Advance system with Cu-K$\alpha$ radiation) was used to determine the phase purity of PBS thin-films. The stoichiometry of PBS thin films was confirmed using energy dispersive spectroscopy in a scanning electron microscope from JEOL. The surface morphology and average film thickness were measured using an atomic force microscope (AFM) from Asylum Research (model:MFP3D).  The longitudinal and Hall resistances were measured in a Quantum Design Physical Property Measurement System (PPMS-ECII) equipped with a $9$~T magnet.

  \begin{figure}
	\includegraphics[width=1.2\linewidth, height=0.35\textheight]{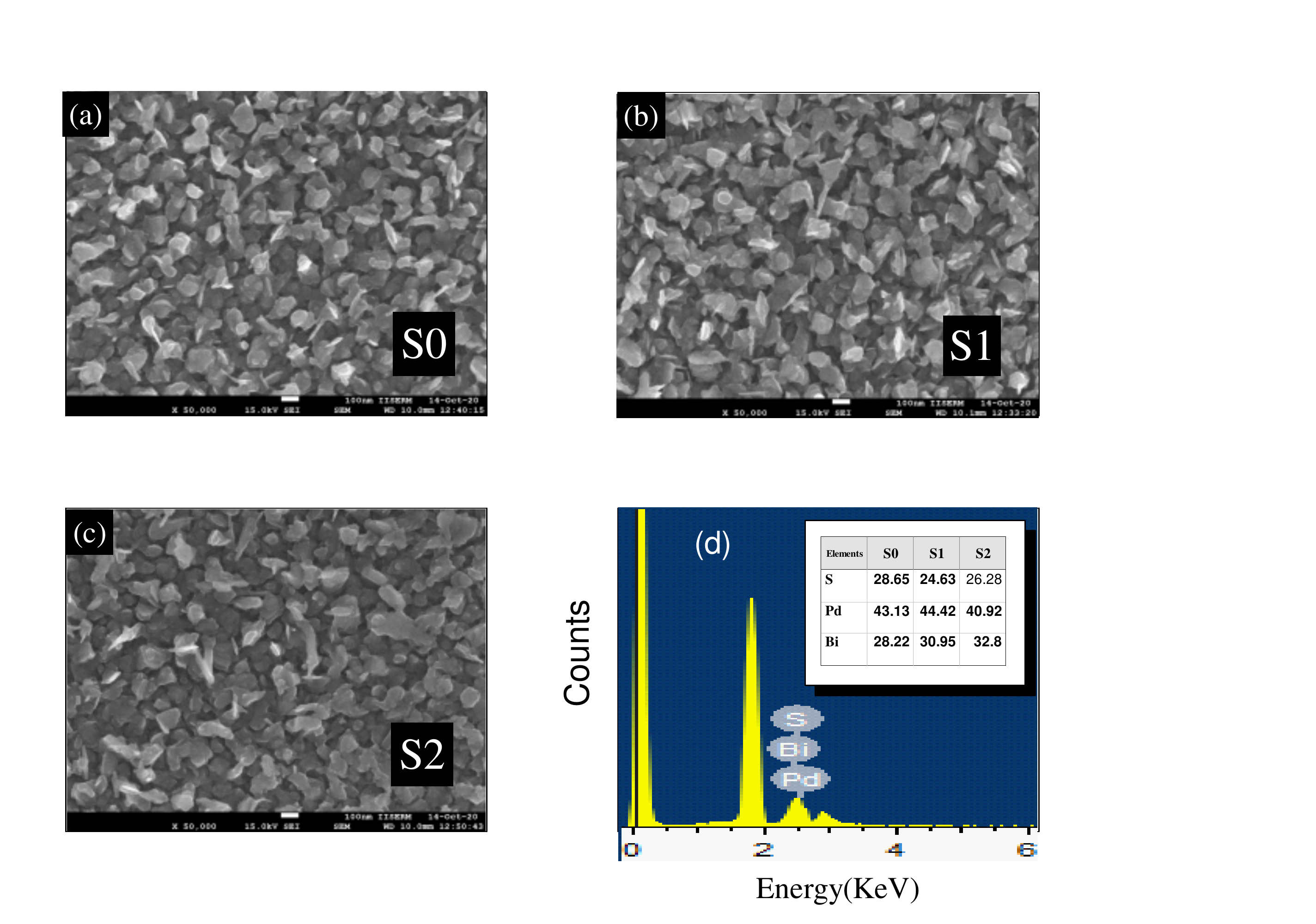}
	\caption{(a-c) Scanning electron microscope images of the three PBS thin films used in this study. (d) The energy dispersive spectroscopy spectrum.  Inset shows a Table giving the measured atomic ratios of elements in the three films.}
	\label{fig-sem}
\end{figure}

\section{Results and Discussion}
Figure~\ref{fig-XRD} shows the x-ray diffraction (XRD) pattern recorded at $T= 300$~K for the as-grown PBS thin film. The XRD pattern confirmed that our thin film crystallizes in the expected cubic crystal structure with space group I2$_{1}$3.  A Rietveld refinement of the XRD pattern gave the lattice parameters $a = b = c = 8.47$~\AA, which are in agreement with values reported previously for bulk PBS. \cite{Roy,Takeshi,Maria}  
Figure~\ref{fig-afm}(a-c) shows the atomic force microscope (AFM) topography image of PBS  thin films.  These images indicate the polycrystalline granular growth of the films.  We also determined the thickness of the film to be $93$~nm as shown in Fig.~\ref{fig-afm}(d). Figure~\ref{fig-sem}(a-c) shows SEM images of all samples. From the AFM and SEM images, it is concluded that PBS thin films are polycrystalline in nature. Figure~\ref{fig-sem}(d) shows energy dispersive X-ray spectroscopy data, where peaks of Bismuth, Palladium, and Sulfur are observed. The atomic ratio of these elements, shown in the Table in the inset, confirms the stoichiometry of the thin films.  Therefore we conclude that we have successfully grown the first polycrystalline thin films of Pd$_{3}$Bi$_{2}$S$_{2}$.

\begin{figure}
	\centering
	\includegraphics[width=0.8\linewidth, height=0.25\textheight]{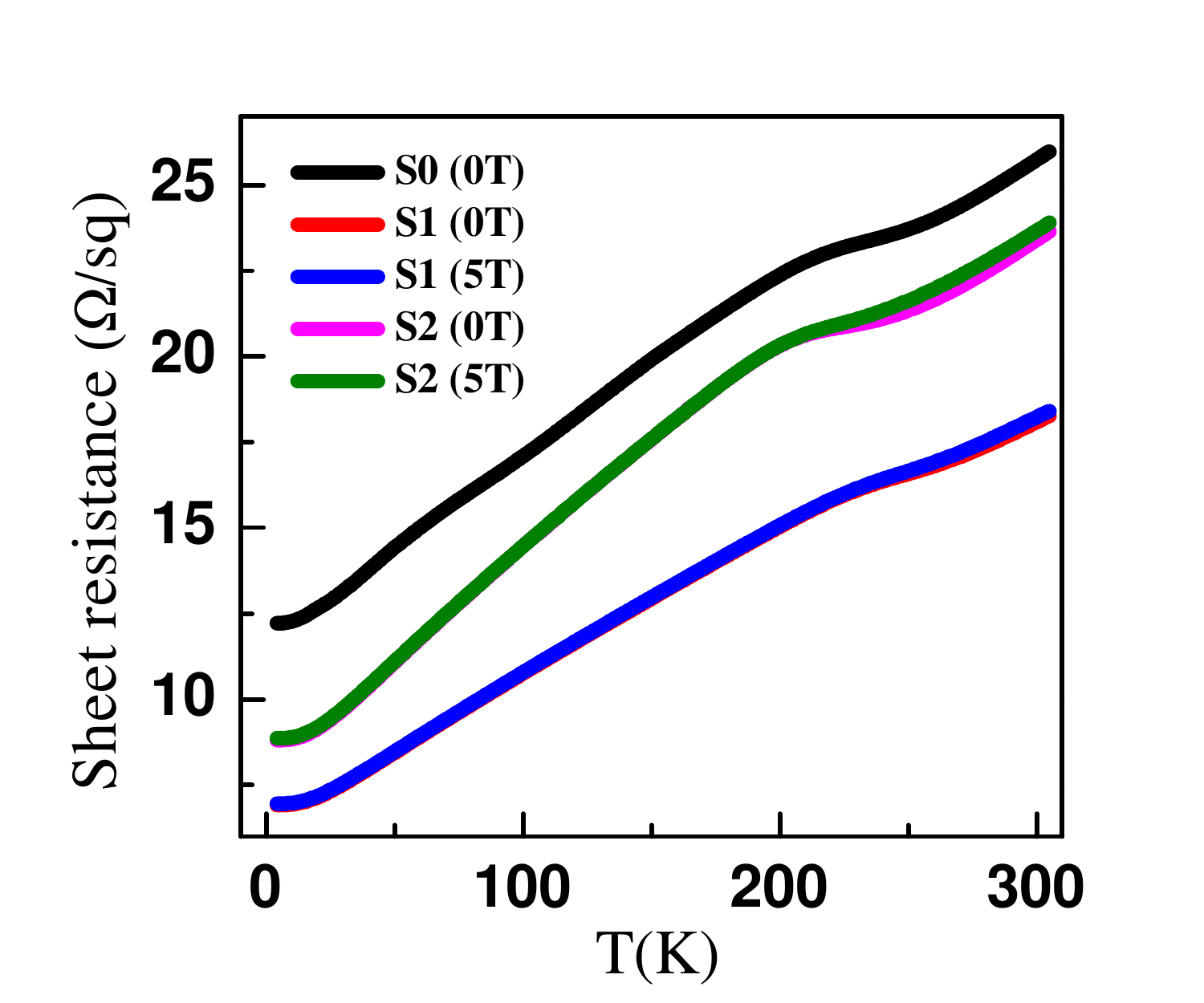}
	\caption{Sheet resistance vs temperature at $B = 0,5$~T showing the metallic behavior of S0-S2 films.}
	\label{fig-rt}
\end{figure}

Figure~\ref{fig-rt} shows the variation of sheet resistance $R_S$ with temperature for PBS (S0-S2) thin films in magnetic fields $B = 0, 5$~T applied perpendicular to the film.  The temperature dependence of $R_S$  reveals the metallic behavior of PBS thin fims with a large residual resistance $R_{o}$.  This metallic behaviour is consistent with previous reports on single crystals.\cite{Roy}  We also observe a slope change in the sheet resistance close to $220$~K seen in all films, which was not observed in bulk samples.\cite{Takeshi,Maria,Roy}  The origin of this slope change is not understood at present but we speculate that the connectivity between grains is changing with temperature, leading to a change in the conductivity of the films. To try to analyze how annealing affects the resistance of different films, we compare the residual resistivity ratio (RRR) and the residual resistivity $R_{o}$ in Table~\ref{Table1}.  From Fig.~\ref{fig-rt} and Table~\ref{Table1} we see that the RRR increases for the annealed films suggesting an improved quality arising most likely from the improved connectivity between grains.  However, the $R_{o}$ varies non-monotonically, suggesting that the optimal annealing temperature for the PBS films may be $260$ at which the S1 films were annealed.

\begin{table}[t]
	\caption{Parameters for Pd$_{3}$Bi$_{2}$S$_{2}$ thin films. $T_{\rm{ann}}$ is the post-annealing temperature, $R_o$ is the sheet resistance at the lowest temperature, RRR is the residual resistivity ratio, $n(p)$ is the electron(hole) density, $\mu_e$($\mu_h$) is the electron(hole) mobility, L$_\phi$ is the phase coherence length, and $\alpha$ is a parameter in the HLN theory which depends on the number of cunduction channels.}
	\setlength{\tabcolsep}{4.25pt} 
	\renewcommand{\arraystretch}{1.4}
	\begin{tabular}{ |c c c c|}
		\hline
		 & S0
		& S1 & S2   \\
		\hline 
		T$_{\rm{ann}}$($^{\circ}$C) & not annealed & 260 & 300   \\
		$R_{o}$($\Omega$) & 12.2 & 6.91 & 8.81  \\ 
		RRR & 2.1 & 2.6 & 2.7  \\ 
		n/p($\times$ 10$^{21}$/cm$^{3}$)& 0.62/1.34 & 0.92/2.26 & 0.83/1.90  \\
		$\mu$$_{e}$/$\mu$$_{h}$($\times$ 10$^{2}$cm$^{2}$/Vs)& 1.73/1.22 & 1.46/.96 & 1.59/1.1 \\
	    L$_{\phi}(nm)$	& 206 & 214 & 205 \\ 
		$\alpha$ & 0.28 & 0.3061 & 0.326  \\ 
\hline
	\end{tabular}
	\label{Table1}
\end{table}

\begin{figure}
	\includegraphics[width=1.1\linewidth, height=0.35\textheight]{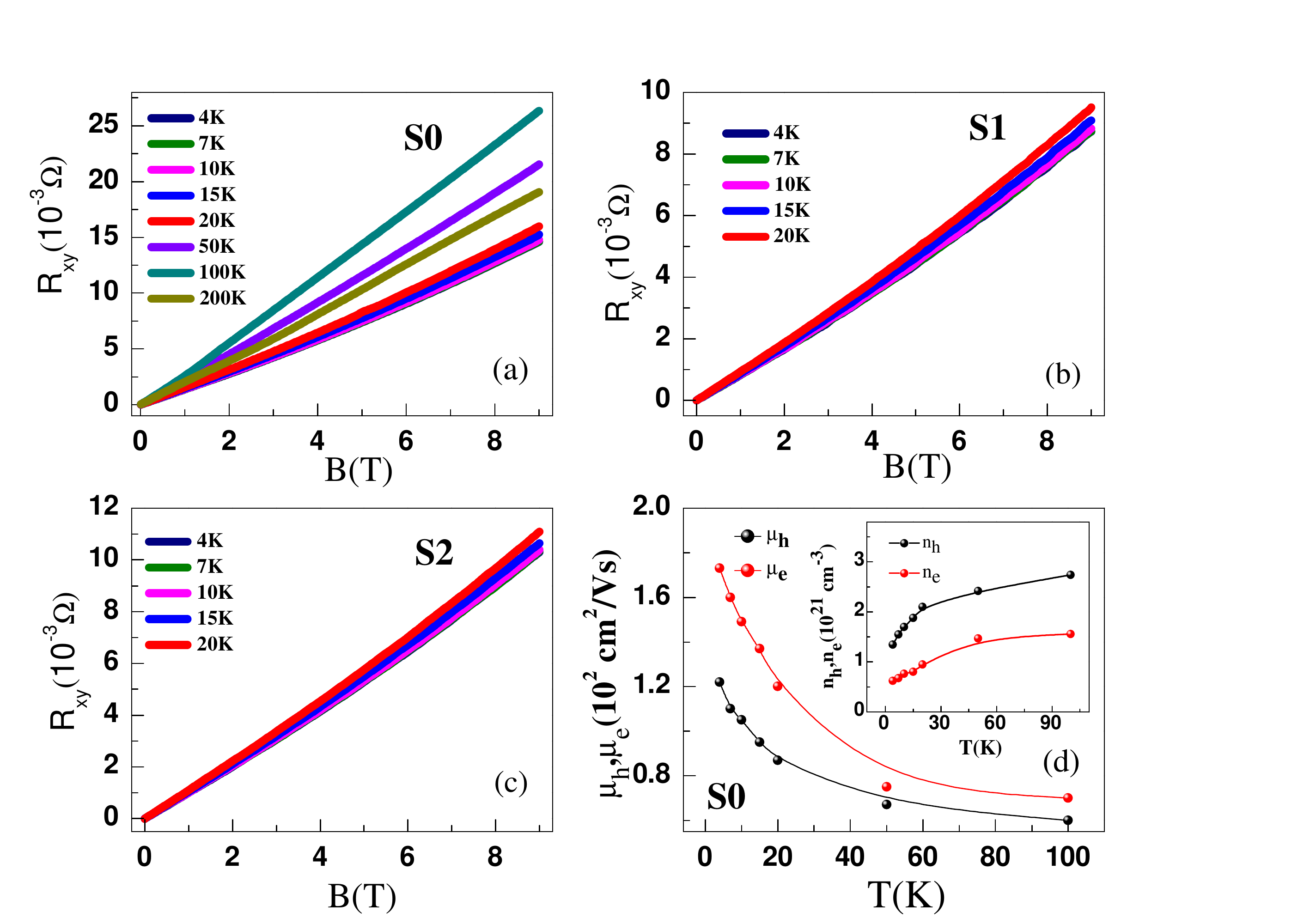}
	\caption{(a-c) Hall resistance (R$_{xy}$) vs magnetic field (B) at various temperatures for S0-S2 thin films of PBS.(d) Variation of mobility with temperature for S0, Inset shows the variation of carrier density with temperature. The solid curves through the data are guides to the eye. }
	\label{fig-hall}
\end{figure}

Figure~\ref{fig-hall}(a-c) shows the variation of Hall resistivity (R$_{xy}$) with magnetic field (B) at various temperatures for S0-S2 films. At high temperature, Hall resistance remains linear and positive with magnetic field for S0 as can be seen from Fig.~\ref{fig-hall}a.  At lower temperatures the Hall resistance becomes non-linear.  A similar non-linear Hall resistance is also observed at low temperatures for the S1 and S2 films as can be seen from Fig.~\ref{fig-hall}(b,c).  A non-linear Hall resistance implies the presence of more than one type of charge carriers in PBS. To extract the carrier concentration and mobility, we fit the Hall data for all films to a two-band model given by the expression\cite{Namrata,Hurd}

\begin{equation}\label{key}
R_{xy}(B) = \dfrac{B}{e}\dfrac{(n_{h}\mu_{h}^{2}- n_{e}\mu_{e}^{2})+(n_{h}-n_{e})\mu_{h}^{2}\mu_{e}^{2}B^{2}}{(n_{h}\mu_{h}+n_{e}\mu_{e})^{2}+(n_{h}-n_{e})^{2}\mu_{h}^{2}\mu_{e}^{2}B^{2}}
\end{equation}

where $e$ is the charge of an electron and $B$ is the magnetic field. The $n$ and $\mu$ are the carrier density and mobility, respectively. The subscript e, h denotes electrons and holes, respectively. The $n$ and $\mu$ estimated from the two band fit to the Hall data, are given in Table~\ref{Table1}. Fig.~\ref{fig-hall}(d) shows the temperature dependence of mobility and carrier density for S0.  The behaviour for S1 and S2 were found to be similar. The electron and hole carrier density is estimated to be of order 10$^{21}$ cm$^{-3}$, which is close to the value reported previously in  PBS single crystals.  However, the mobility for the PBS films comes out to be of the order 10$^{2}$ cm$^{2}$/Vs which is smaller than reported in single crystals.\cite{Roy}  From the extracted parameters we can rule out the possibility of charge carrier compensation and we also conclude that holes are the dominant charge carriers.   The extracted parameters for all samples are shown in Table~\ref{Table1}.

\begin{figure}
	\centering
	\includegraphics[width=0.8\linewidth, height=0.4\textheight]{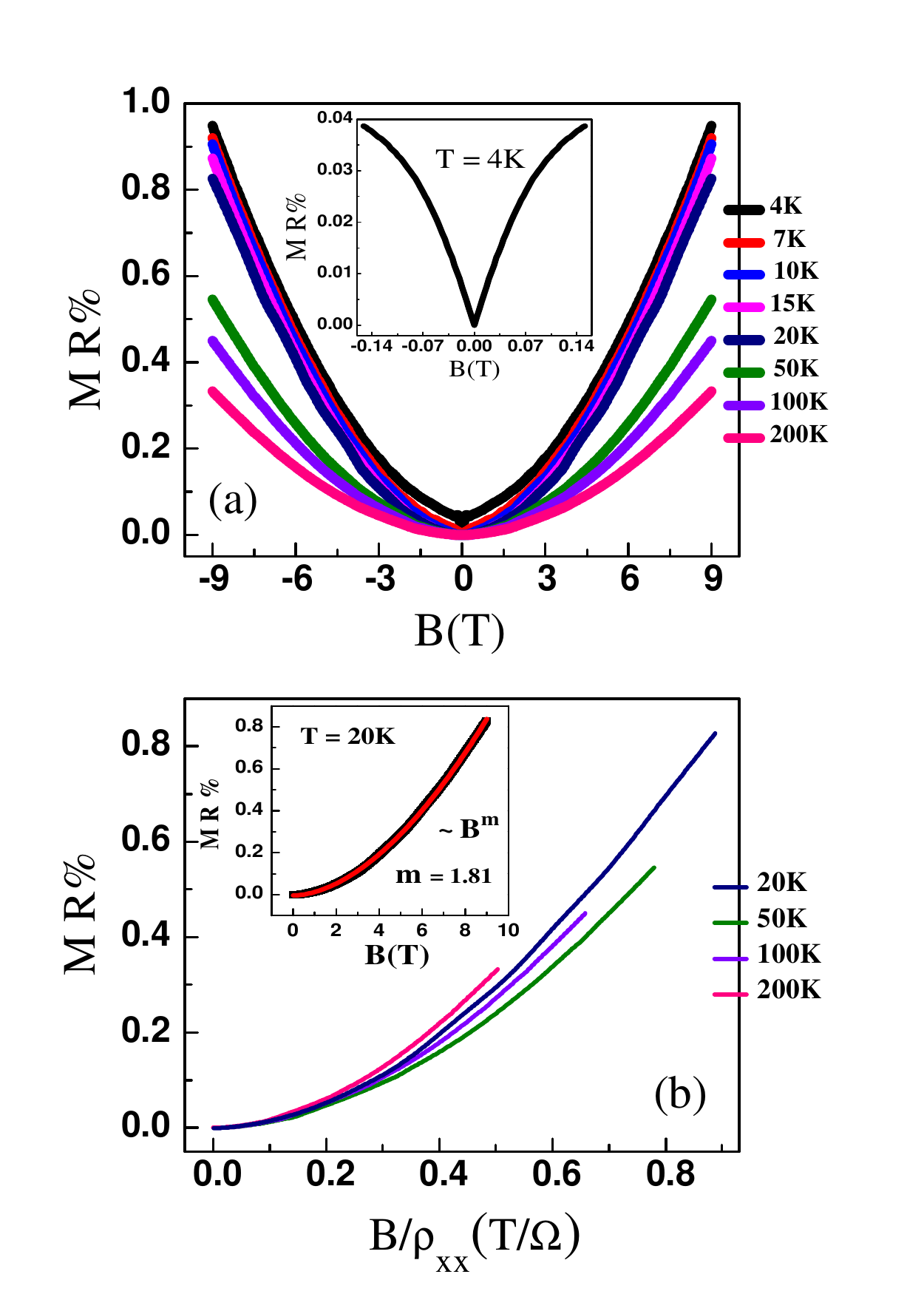}
	\caption{(a) The percentage magneto-resistance MR vs magnetic field $B$ at various temperatures for S0 film.  The inset shows the data at low magnetic fields at $T= 4$K showing the WAL effect. (b) Violation of Kohler's rule indicating the presence of multiple scattering mechanism. The inset shows the power-law $B^m$ fitting of the MR data at $T= 20$~K with $m = 1.81$.}
	\label{fig-mr}
\end{figure}

\begin{figure*}
	\centering
	\includegraphics[width=.7\linewidth, height=0.4\textheight]{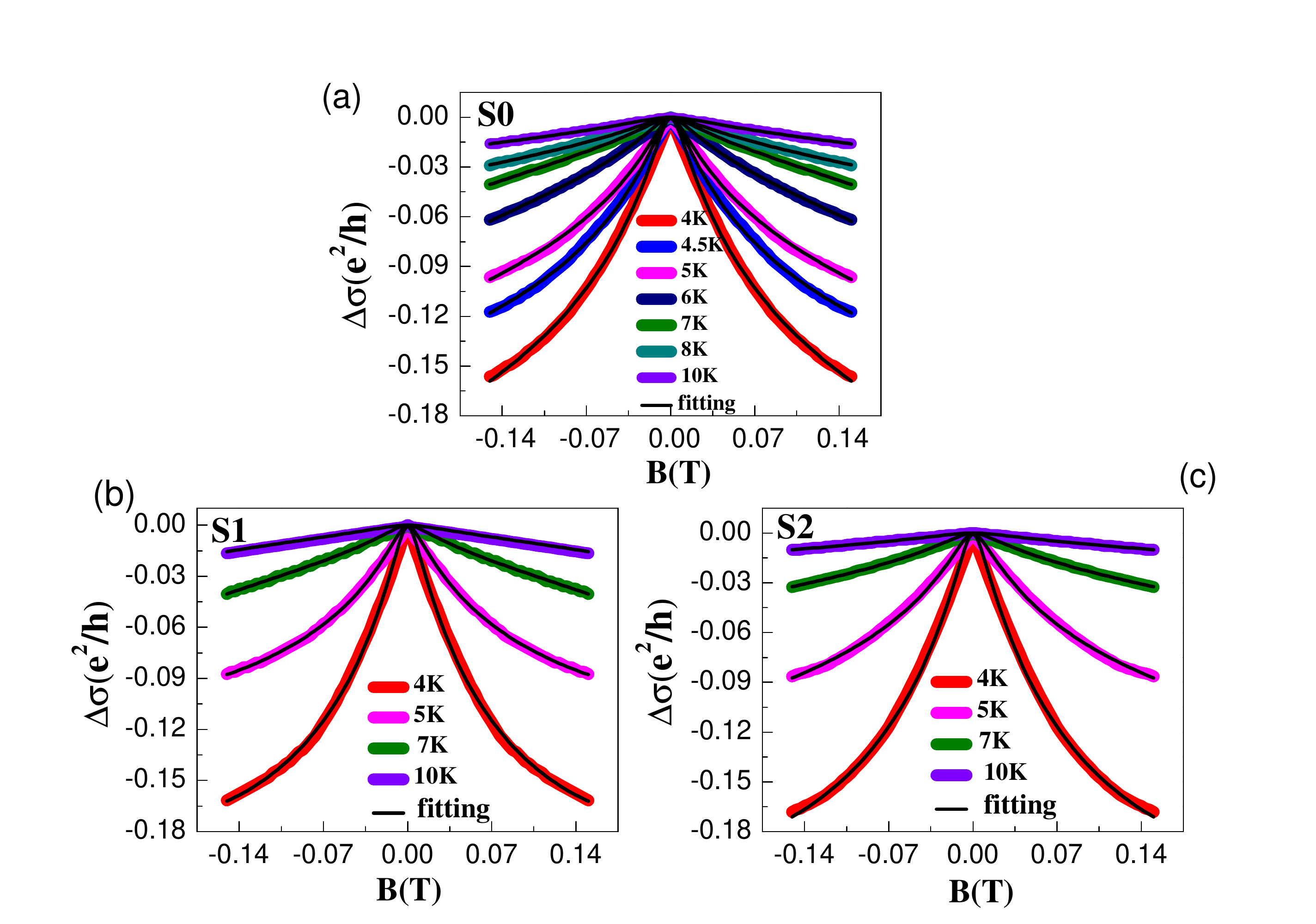}
	\caption{The magneto-conductance ($\triangle$$\sigma$) versus field $B$ at various temperatures for (a) as-grown film S0 and for post annealed films (b) S1 and (c) S2. Solid curves through the data  are fits to the HLN equation.}
	\label{fig-WAL}
\end{figure*}

\begin{figure*}
	\centering
	\includegraphics[width=.7\linewidth, height=0.2\textheight]{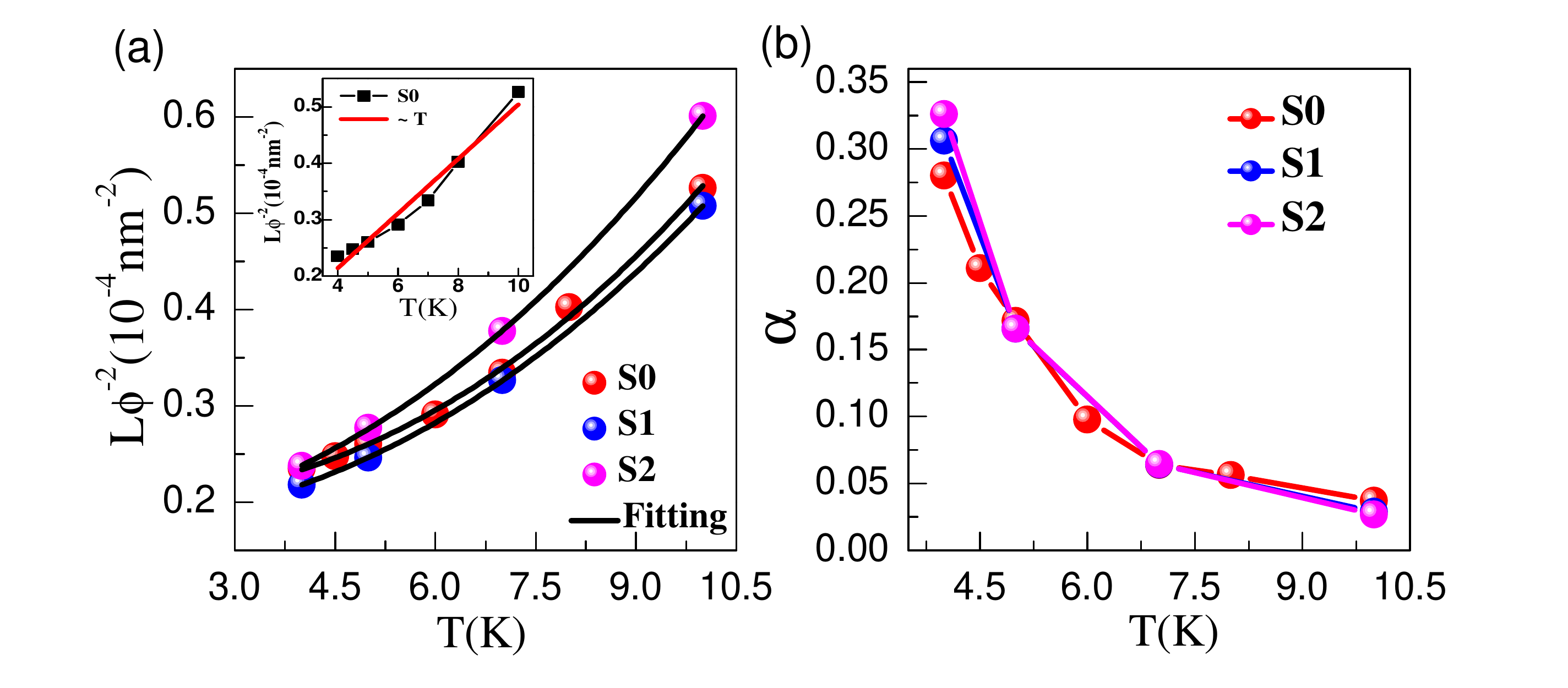}
	\caption{(a) Variation of L$_{\phi}$ as a function of temperature, revealing the contribution of different scattering mechanisms. Inset shows the failure of a linear in $T$ fitting.  (b) Temperature dependence of $\alpha$. }
	\label{fig-lphi}
\end{figure*}

Figure~\ref{fig-mr}(a) shows the percentage magnetoresistance MR\% at various temperatures where 
\begin{equation}\label{key}$$\centering
MR\% = $\dfrac{R(B) - R(0)}{R(0)}$ $\times$ 100~.
$$\end{equation}

Here $R(B)$ is the sheet resistance in a magnetic field $B$.  The inset in Figure~\ref{fig-mr}(a) shows the MR data at low fields highlighting the cusp-like behaviour at low magnetic fields. This cusp-like behavior under low field is a signature of weak anti-localization (WAL).
\cite{Bergmann,Hikami,Fang,Lee} The low value of the magneto-resistance in our films compared to the previous report on single crystals\cite{Roy} is mainly attributed to the low carrier mobility in PBS thin films, as observed from Hall data.  According to Kohler’s rule, for materials with one dominant scattering mechanism, the MR can be represented as a universal function of the quantity $B\tau$, where $\tau$ is the time between scattering events of conduction electrons which is inversely proportional to the zero field resistivity $\rho$$_{o}$.\cite{Ziman}  We therefore expect that the MR is a universal function of the ratio $B/\rho_o$.
Figure~\ref{fig-mr}(b) shows the MR data for the S0 film plotted as a function of B/$\rho_{o}$.  If Kohler's rule was obeyed, the MR vs B/$\rho_{o}$ data would collapse onto a single curve. Figure~\ref{fig-mr}(b) shows that the Kohler's rule is violated for PBS films.  This indicates the presence of more than one dominating scattering mechanism.
 
We next present the WAL analysis.  However, the analysis depends on the dimensionality of the system. For thin films, we must estimate if the mean free path is greater than the film's thickness.  In our case, we have calculated the mean free path using the 2D formula, $l_{e} = \dfrac{\hbar}{e}\sqrt{2\pi n_{e}}\mu_{e}$, where n$_{e}$ and $\mu_{e}$ are the 2D Hall carrier density and mobility. We estimate $l_{e} = 126$~nm for S0, which is larger than the film thickness ($93$ñm). For Quantum interference (QI) effects, the relevant length scale is the phase coherence length $L_{\phi} = \sqrt{D\tau}$, where D is the diffusion constant and $\tau$ is the phase coherence time.  Also, the criterion for 2D nature of thin films is $L_\phi>$ thickness of film.\cite{Anderson,Ovadyahu} In our case, the value of L$_{\phi}$ is found to be greater than the thickness of thin-film (based on the analysis presented below), indicating the 2D nature of our thin films.

Figure~\ref{fig-WAL}(a-c) shows the conductance at various temperatures in the low magnetic field range $|B| \leq 0.15$~T for all films.The two dimensional conductance was found using
$\triangle$$\sigma$ = $\sigma$(B)-$\sigma$(0) where $\sigma$(B) = (L/W) (1/R$_{xx}$), and L and W are the length and width of film respectively. For two-dimensional systems, Hikami-Larkin-Nagaoka (HLN) equation can be used to model the effect of localization.\cite{Hikami,Fang} This theory involves the contribution of quantum effects from three different mechanisms, namely, spin-orbit coupling (SOC), elastic scattering, and electron phase coherence. In the limit of high SOC, Hikami-Larkin-Nagaoka (HLN) can be written as
\begin{equation}\label{key}$$
\centering
$\triangle$ $\sigma$(B)= -$\alpha$$\dfrac{e^{2}}{\pi h}$\bigg[$\psi$\bigg($\dfrac{1}{2}$+$\dfrac{B_{\phi}}{B}$ \bigg)- ln \bigg($\dfrac{B_{\phi}}{B}$\bigg)\bigg]
$$\end{equation}
where $\psi$
is the digamma function, e is the electron charge, h is the Planck constant, B$_{\phi}$ = $\hbar^{2}$/(4eL$_{\phi}^{2}$) is the characteristic field associated with phase coherence length L$_{\phi}$.  At higher magnetic fields, we must include the classical MR contribution parabolic in $B$. One can therefore write down an expression which contains the contribution from both classical and quantum parts as\cite{Assaf}

\begin{equation}$$
\centering
$\triangle$ $\sigma$(B)= -$\alpha$$\dfrac{e^{2}}{\pi h}$\bigg[$\psi$\bigg($\dfrac{1}{2}$+$\dfrac{B_{\phi}}{B}$ \bigg)- ln \bigg($\dfrac{B_{\phi}}{B}$\bigg)\bigg]+ cB$^{2}$.
\label{HLN}
$$\end{equation}
The parameter $\alpha$ gives the number of conduction channels contributing to the transport. According to the HLN theory, parameter $\alpha$ takes value 1/2 and -1 for weak anti-localization (WAL) and localization (WL), respectively. However, $\alpha$  has often been observed to deviate from $1/2$ in the case when more than one conducting channels are available.\cite{Zhao} 

In our material Pd$_{3}$Bi$_{2}$S$_{2}$, the spin orbit coupling (SOC) is also expected to be high. We fit the experimental data by Eq.~5 with fitting parameters $\alpha$ and L$_{\phi}$.  The temperature dependence of the obtained fitting parameters are shown in Fig.~\ref{fig-lphi}. The extracted value of $\alpha = 0.28$ at $T= 4$~K is smaller than the theoretical value $0.5$ expected for a single conduction channel. This indicates the presence of more than one conduction channel. For example, the presence of Topologically trivial bands at the Fermi level could contribute to the conductivity in addition to the Topological electrons.  Figure~\ref{fig-lphi}(b) shows that $\alpha$ reduces from $0.28$ at $4$~K to $0.04$ at $10$~K, consistent with the trend previously observed for other Topological materials like Cd$_3$As$_2$.\cite{Zhao} 

Figure~\ref{fig-lphi}(a) shows the temperature dependence of phase coherence length L$_{\phi}$ plotted as $1/L_{\phi}^2$ versus $T$ for as-grown and post annealed films. The L$_{\phi}$ decreases from $210$~nm at $4$~K to $138$~nm at $10$~K\@.  The fact that L$_\phi$ comes out to be larger than the film thickness validates our use of 2D WAL analysis for our PBS films.  The Nyquist theory, which considers electron-electron scaterring, predicts that L$_{\phi}$$\propto$ T$^{-1/2}$ for 2D systems.\cite{Wu} The inset of Fig.~\ref{fig-lphi}(a) shows that the L$_{\phi}$ data for PBS doesn't follow the T$^{-1/2}$ dependence. This indicates that multiple scattering mechanisms with different temperature dependences could be involved in the dephasing of the electron's phase in PBS. We try to analyze the temperature dependence of L$_\phi$ using the simple equation\cite{Bird}

\begin{equation}\label{key}\centering
\dfrac{1}{L_{\phi}^{2}} =\dfrac{1}{L_{\phi o}^{2}} + A_{ee}T + B_{ep} T^p
\end{equation}

where $L_{\phi o}$ represents the zero temperature dephasing length, and A$_{ee}T$ and B$_{ep}$ T$^{p}$ represent the contributions from electron-electron and electron-phonon interactions, respectively.  The value of $p$ for electron-phonon interaction in 2D materials can vary between $2$ to $3$ depending on the effective dimensionality and disorder in the film.\cite{Sergeev,Reizer,Lawrence,Breznay}  We obtained an excellent fit of the temperature dependent L$_{\phi}$ data for all films by the Eq. 6 with $ p \approx 2.4$ as shown by the solid curves through the data in Fig.~\ref{fig-lphi}(a) similar to values found on disordered films of GeSb$_2$Te$_4$. \cite{Breznay}
 
\section{Conclusion}
We report the first synthesis of thin films ($93$~nm) of the novel Topological material Pd$_3$Bi$_2$S$_2$ grown on Si (111) substrate by pulsed laser deposition. The resistance measurements on PBS thin films indicate a disordered metallic system.  The magneto-resistance data show deviations from Koehler's rule suggesting the presence of multiple scattering mechanisms.  We also find Quantum interference effects in the observation of the weak anti-localization effect at low magnetic fields. These results are satisfactorily analyzed in terms of Hikami-Larkin-Nagaoka theory in the presence of spin-orbit coupling. It was found that the coefficient $\alpha$ deviates from the value $0.5$ expected for 2D systems with a single conduction channel.  This indicates the contribution from several conducting channels in the electron transport of PBS films. Dependence of the dephasing length L$_{\phi}$ on temperature is also anomalous. We found that this behaviour can be understood by including both the Nyquist electron-electron scattering as well as electron-phonon scattering as the phase relaxation mechanism in PBS films.

\section{acknowledgment}
We thank the x-ray and SEM facilities at IISER Mohali.

\end{document}